\documentclass[12pt]{article}
\usepackage{lscape,slashed}
\usepackage{multirow}
\usepackage{color}
\usepackage{bm}
\usepackage{comment}
\usepackage{amssymb,amsmath,yhmath,mathrsfs,graphicx}
\usepackage{amstext}
\usepackage{amscd}
\usepackage{graphics}
\usepackage{epsfig}
\usepackage{shadow}
\usepackage[all]{xy}
\usepackage{hyperref}

\def\m{\mu}
\def\n{\nu}

\def\be{\begin{equation}}
\def\ee{\end{equation}}
\def\bes{\begin{equation*}}
\def\ees{\end{equation*}}
\def\beq{\begin{equation}}
\def\eeq{\end{equation}}
\def\bea{\begin{eqnarray}}
\def\eea{\end{eqnarray}}
\def\beas{\begin{eqnarray*}}
\def\eeas{\end{eqnarray*}}

\def\nn{\nonumber}

\def\sideremark#1{\ifvmode\leavevmode\fi\vadjust{\vbox to0pt{\vss
 \hbox to 0pt{\hskip\hsize\hskip1em
 \vbox{\hsize2cm\tiny\raggedright\pretolerance10000
  \noindent #1\hfill}\hss}\vbox to8pt{\vfil}\vss}}}

\begin{document}
\thispagestyle{empty}

\setcounter{footnote}{0}

\vspace{-50mm}
\begin{flushright}\tt CALT-TH 2015-017,\! BRX-TH-6294 \end{flushright}

\begin{center}

{\Large
 {\bf  Propagation peculiarities  of mean field massive gravity}\\[4mm]

 {\sc \small
    S.~Deser$^{\rm a}$, A.~Waldron$^{\rm b}$ and G. Zahariade$^{\rm c}$}\\[3mm]

{\em\small
                      ~\ ${}^\mathfrak{\rm a}$
                       Walter Burke Institute for Theoretical Physics, California Institute of Technology, Pasadena, CA 91125; 
     Physics Department, Brandeis University, Waltham, MA 02454.\\
{\tt deser@brandeis.edu}\\[2mm]
          
           ~${}^{\rm b}\!$
            Department of Mathematics\
            University of California,
            Davis CA 95616, USA\\
            {\tt wally@math.ucdavis.edu}

           ~${}^{\rm c}\!$
            Department of Physics\
            University of California,
            Davis CA 95616, USA\\
            {\tt  zahariad@ucdavis.edu}\\[2mm]

            }
}

\bigskip

{\sc Abstract}\\[1mm]

\end{center}

\newcommand{\mGR}{$\overline {\text m\hspace{-.1mm}}$GR\ }

\noindent
Massive gravity (mGR) describes a dynamical ``metric'' on a fiducial, background one. We investigate fluctuations 
of the dynamics about mGR solutions, that is about its ``mean field theory''. Analyzing mean field massive gravity (\mGR\!\!) propagation characteristics is not only equivalent to studying those of the full non-linear theory,
but also in direct correspondence with earlier analyses of charged higher spin systems, the oldest example being the charged, massive spin~$3/2$ Rarita--Schwinger (RS) theory.
The fiducial and mGR mean field background metrics in the \mGR model  correspond to the RS Minkowski metric and external EM field. The common implications in both systems are that hyperbolicity holds only in a weak background-mean-field limit, immediately ruling both theories out as fundamental theories. Although both can still be considered as
predictive effective models in the weak regime, their lower helicities  exhibit superluminal behavior: lower helicity gravitons are superluminal as  compared to photons propagating on either the fiducial or background metric.  This ``crystal-like'' phenomenon of differing helicities having differing propagation speeds in both \mGR and mGR  is a {\it peculiar} feature of these models.

\newpage



\section{Introduction}

Consistency is a powerful tool for studying field theories. Already classically, there are stringent conditions that are extremely difficult to fulfill for systems with spin~$s>1$, the most important exception being ($s=2$, $m=0$)  general relativity (GR).
Key consistency requirements are
\begin{enumerate}
\item[(i)] Correct degree of freedom (DoF) counts.
\item[(ii)] Non-ghost kinetic terms.
\item[(iii)] Predictability.
\item[(iv)] (Sub)luminal propagation.
\end{enumerate}
Requirements (i) and (ii) are closely related (as are  (iii) and (iv)). Models whose constraints do not single out the correct propagating DoF suffer from relatively ghost kinetic terms: the relevant example here  is the sixth ghost excitation that plagues generic massive gravity (mGR) theories~\cite{BD}.
The discovery that a class of mGR models 
satisfied requirements (i) and~(ii)
generated a revival of interest in massive spin~2 theories~\cite{dRGT,HR,DMZ,DW,DSW,DSWZ} even though failure of the  propagation requirements (iii) and~(iv) were long known to bedevil higher spin theories~\cite{VZ,KoS}. 

The predictability requirement is that initial data can be propagated to the future of spacetime hypersurfaces. In PDE terms, this means that the underlying equations must be hyperbolic~\cite{CourantHilbert}.
The final requirement, that signals cannot propagate faster than light, can be imposed once the hyperbolicity requirement is satisfied. The classic example of a model
that obeys requirements (i) and (ii) as well as (iii) but only in a weak field region, is the charged, massive,~$s=3/2$ RS theory.
Curiously enough, the propagation problems
of this model were first discovered in a quantum setting by Johnson and Sudarshan~\cite{JS} who studied the model's canonical field commutators (this is easy to understand in retrospect, because field commutators and propagators are  directly related~\cite{BjD}). The first detailed analysis of the model's propagation characteristics was carried out by Velo and Zwanziger; our aim is to reproduce their RS results in \mGR\!, so we quote their 1971 abstract verbatim~\cite{VZ}:
\begin{quote}
The Rarita--Schwinger equation in an external electromagnetic potential is shown to be equivalent to a hyperbolic system of partial differential equations supplemented by initial conditions. The wave fronts of the classical solutions are calculated and are found to propagate faster than light. Nevertheless, for sufficiently weak external potentials, a consistent quantum mechanics and quantum field theory may be established. These, however, violate the postulates of special relativity.
\end{quote}
In previous works we and other authors have shown that similar  conclusions hold for the full non-linear mGR models~\cite{DW,DSW,Izumi,DeserOng,DSWZ,Gr}. These investigations rely on 
the method of characteristics, which amounts to studying leading kinetic terms 
and is thus essentially equivalent to an analysis of linear fluctuations around a mean field background. Since this mean field massive
gravity (\mGR\!\!) fluctuation model 
depends both on a background and a fiducial metric, it is in direct correspondence with the charged RS model.
Hence, without any computation at all, one can readily predict that: (a) mGR loses hyperbolicity in some strong field regime
and (b) in the weak field hyperbolic regime where predictability is restored, lower helicity modes have  propagation characteristics differing from maximal helicity~$\pm 2$; thus superluminality with respect to (luminal) photons is inevitable. 
Apart from confirming earlier conclusions in a very simple setting, our results give a precise description of  mGR's effective, weak field, regime.

\section{Massive Gravity}

At its genesis, the first known non-linear mGR model of~\cite{Zumino} was originally formulated in terms of dynamical and fiducial vierbeine~$e^m$ and~$f^m$. It took some forty years for researchers---independently in an effective field theory-inspired metric formulation---to discover that this model was one of a three-parameter
family~\cite{dRGT} that avoided the sixth, ghost-like excitation of~\cite{BD}.
The action describing these fiducial mGR models is given by\footnote{Here $d$ is the exterior derivative and the dynamical vierbiene and spin connection $(e,\omega)$, are one-forms. We suppress wedge products unless necessary for clarity.}
\bes
\begin{split}
{S}_{\rm mGR}[e,\omega;f]=-\, \int \epsilon_{mnrs}e^m &\left\{ \ \,   \frac{1}{4}\,   e^n \left[d\omega^{rs} + \omega^r{}_t{{}}\omega^{ts}\right]\phantom{\frac{\beta}{4}}\right.\\
&\ \, -\left.
m^2  \left[\frac{\beta_{0}}{4}e^n{{}} e^r{{}} e^s 
 +\frac{\beta_{1}}{3}e^n{{}} e^r{{}} f^s+\frac{\beta_{2}}{2}e^n{{}} f^r{{}} f^s+\beta_{3}f^n{{}} f^r{{}} f^s\right]\right\}\, .
\end{split}
\ees
The parameter $\beta_0$ governs a standard cosmological term; this is required to obtain the Fierz--Pauli (FP) linearized limit when both the fiducial and mGR backgrounds are Minkowski. When both the fiducial and mGR  backgrounds are Einstein  with cosmological constant $\bar \Lambda$, the model's parameters must obey 
$\frac{\bar \Lambda}{3!} = m^2 \left(\beta_0+\beta_1+\beta_2+\beta_3\right)$
and the linearized theory is~FP  with mass
$
m_{\rm FP}^2:=m^2(\beta_1+2\beta_2+3\beta_3)$.

Varying the model's dynamical fields~$(e^m, \omega^{mn})$ gives equations  of motion
\bea\label{mGReoms}
\nabla e^{m}\approx 0\approx G_{m}-m^{2}\, t_m
\, ,
\eea
where
$t_m:=\epsilon_{mnrs}\big[\beta_0 e^n e^r e^s +
\beta_1 e^n e^r f^s +
\beta_2 e^n f^r f^s +
\beta_3 f^n f^r f^s  \big]
$. 
Also, 
 the Einstein three-form 
is defined by $G_{m}:=\frac{1}{2}\epsilon_{mnrs}e^{n}{{}} R^{rs}$
and $R^{mn}:=d\omega^{mn}+\omega^{m}{}_{r}{{}}\omega^{rn}$ is the Riemann curvature; $\nabla$ is the connection of $\omega^{mn}$. The forty equations above are subject to thirty constraints that are spelled out in detail in~\cite{DSWZ}. In particular, these include the covariant algebraic relations\footnote{The first of these assumes invertibility of the operator $M^{mn}$ as a map from two-forms to antisymmetric Lorentz tensors; we shall always work in the model's branch where this holds.}
$$
e^m f_m\approx 0 \approx K_{mn} e^m f^n\approx \epsilon_{mnrs} M^{mn}K^{rs}\, ,
$$ 
where the tensor $K^{mn}:=\omega^{mn}-\chi^{mn}$ denotes the contorsion and $M^{mn}:=\beta_1 e^m e^n + 2 \beta_2 e^{[m}f^{n]} + 3 \beta_3 f^m f^n$.

\section{Mean field massive gravity}\label{cca}

Consider mGR propagating in an arbitrary fiducial (pseudo-)Riemannian manifold $(M,\bar g_{\m\n})$ with corresponding vierbeine and spin connections $(f^m,\chi^{mn})$. Now let $(e^m,\omega^{mn})$ be a solution
to the mGR equations of motion~\eqref{mGReoms}. We wish to study fluctuations $(\varepsilon^m,\lambda^{mn})$ about this configuration:
$$
\tilde e^m = e^m+\varepsilon^m\, ,\quad
\tilde \omega^{mn}=\omega^{mn}+\lambda^{mn}\, . 
$$
The action governing these is the quadratic part of $S_{\rm mGR}[\tilde e,\tilde \omega;f]-S_{\rm mGR}[e,\omega;f]$, namely
\begin{equation*}
\begin{split}
S[h,\lambda;e,f]:=-\frac12 \int \epsilon_{mnrs}\Big[e^m\varepsilon^n\nabla \lambda^{rs} &+\, \, \frac12\, \, \big(e^m e^n \lambda^r{}_t \lambda^{ts}+R^{mn} \varepsilon^r \varepsilon^s\big)\\ &-m^2\big(3\beta_0 e^m e^n \varepsilon^r \varepsilon^s
+2\beta_1 e^m f^n \varepsilon^r \varepsilon^s
+\beta_2 f^m f^n \varepsilon^r \varepsilon^s\big)
\Big]\, .
\end{split}
\end{equation*}
The mean field model is a theory of forty dynamical fields~$(\varepsilon^m,\lambda^{mn})$.
In the above, $\nabla$ is the Levi-Civita connection of $e^m$, and $R^{mn}$ its Riemann tensor; we stress that henceforth the fiducial field~$\big(f^m,\chi^{mn}(f),\bar g_{\mu\nu}(f)\big)$ and mGR background fields~$\big(e^m, \omega^{mn}(e), g_{\mu\nu}(e)\big)$ are {\it non-dynamical}; all index manipulations will be carried out using the mGR background metric and vierbein. 

The \mGR equations of motion are
\begin{eqnarray}\label{mbGR}
{\mathcal T}^m&:=& \nabla \varepsilon^m
+\lambda^{mn} e_n
\approx 0\, ,\nn\\[1mm]
{\mathcal G}_m&:=& \frac12\epsilon_{mnrs}\big[
e^n \nabla \lambda^{rs}+\varepsilon^n R^{rs}\big]
-m^2 \, \tau_m
\approx 0\, ,
\end{eqnarray}
where
$
\tau_m:=\epsilon_{mnrs}\Big[3\beta_0\,  e^n e^r \varepsilon^s + 
2 \beta_1 \, e^n f^r \varepsilon^s + \beta_2 \, f^n f^r \varepsilon^s\Big]
$.

\section{Mean field degrees of freedom}

In principle, since we are describing the linearization of a model whose constraints have been completely analyzed in~\cite{DSWZ}, we know {\it a priori} that \mGR describes five propagating degrees of freedom. However
for completeness and our causality study, we reanalyze its constraints.

The first step is to introduce a putative choice of time coordinate~$t$, 
which for now need not rely in any way on either the fiducial or background metric, and use this to decompose any~$p$-form~$\mbox{\resizebox{2.5mm}{2.85mm}{$\theta$}}$ (with~$p<4$) as
\bes\label{ring}
 \mbox{\resizebox{2.5mm}{2.85mm}{$\theta$}}:= \bm \theta+\mbox{\resizebox{2.5mm}{4.1mm}{$\ring\theta$}}\, ,
\ees
where~$\mbox{\resizebox{2.5mm}{4.1mm}{$\ring\theta$}}\wedge dt=0$.
 Thus~$\bm \theta$ is the purely spatial part of the form~\resizebox{2.5mm}{2.85mm}{$\theta$}. Hence
 for any on-shell relation~${\cal P}\approx 0$ polynomial in $(\nabla,\varepsilon,\lambda)$,
  its spatial~$\bm {\mathcal P}\approx 0$ part   is a constraint because it contains no $t$-derivatives.

Thus we immediately find {\it sixteen} primary constraints:
$$
{\bm {\mathcal T}}^{m}={\bm \nabla}{\bm \varepsilon}^m+{\bm \lambda}^{mn}{\bm e}_n\approx 0\approx
{\bm {\mathcal G}}_{m}=
 \frac12\epsilon_{mnrs}\big[
{\bm e}^n {\bm \nabla \bm \lambda}^{rs}+{\bm \varepsilon}^n {\bm R}^{rs}\big]
-m^2\, {\bm \tau}_m\, .
$$
There are {\it ten} secondary constraints in total:
The first six of these follow from the integrability condition $e^{[m} \nabla {\mathcal T}^{n]}\approx 0$
which yields the so-called {\it symmetry constraint}
$$
e_{[m} \tau_{n]}+\varepsilon_{[m} t_{n]}\approx 0\, .
$$
As mentioned above, we assume that the set of two forms $\{M^{mn}\}$
is a  basis for the space of two-forms, so the symmetry constraint yields
$$
\varepsilon^{m}f_{m}\approx 0\ .
$$
The remaining four secondary constraints come from the covariant curl $\nabla{\mathcal G}_m\approx 0$ and give the {\it vector constraint}
$$
\nabla \tau_m+\lambda_m{}^n t_n\approx 0\, .
$$
Employing the equation of motion ${\mathcal T}^m\approx 0$, this implies
$$
\epsilon_{mnrs} \big[M^{mn} \lambda^{rs} + 2 (\beta_1 e^m +\beta_2 f^m)\, \varepsilon^n K^{rs}\big]\approx 0\, .
$$
Finally there are {\it four} tertiary constraints stemming from covariant curls of the secondaries: the temporal part of the {\it curled symmetry constraint} $K_{mn}\varepsilon^{m}f^{n}+\lambda_{mn}e^{m}f^{n}\approx 0$, {\it i.e.}
$$
\ring K_{mn} \bm\varepsilon^m \bm f^n+\bm K_{mn} \ring \varepsilon^m \bm f^n+\bm K_{mn} \bm \varepsilon^m \ring f^n
+\ring \lambda_{mn} \bm e^m \bm f^n+\bm \lambda_{mn} \ring e^m \bm f^n+\bm \lambda_{mn} \bm e^m \ring f^n
\approx 0\ ,
$$
and the {\it scalar constraint}
\bea
&&\epsilon_{mnrs}\left(\beta_{1}(\varepsilon^{m}e^{t}+e^{m}\varepsilon^{t})-2\beta_{2}\varepsilon^{(m}f^{t)}\right)K^{nr}K^{s}{}_{t}\nn\\[2mm]
&&\quad\quad+ \epsilon_{mnrs}\left(\beta_{1}e^{m}e^{t}-2\beta_{2}e^{(m}f^{t)}-3\beta_{3}f^{m}f^{t}\right)(\lambda^{nr}K^{s}{}_{t}+K^{nr}\lambda^{s}{}_{t})\nn\\[2mm]
&&\quad\quad+\, 2\, m^{2}\, \beta_{1}\varepsilon^{m}t_{m} + 2\, m^{2}\left(\, \beta_{1}e^{m}+2\beta_{2}f^{m}\right)\tau_{m}+3\, \epsilon_{mnrs}\beta_{3}f^{m}f^{n}\nabla\lambda^{rs}\nn\\[2mm]
&&\quad\quad-\, 4\, \beta_{2}\varepsilon^{m}\bar{G}_{m}-2\epsilon_{mnrs}\beta_{1}\varepsilon^{m}e^{n}\bar{R}^{rs}\approx 0\nn\ .
\eea
The  $\nabla\lambda^{rs}$ term seems to indicate that the above display is not a constraint for $\beta_3\neq 0$, however as shown in~\cite{DSWZ},
this quantity (weakly) equals one without time derivatives of fields. In summary, the model describes forty fields subject to thirty constraints and thus propagates five\footnote{Although this conclusion for \mGR is guaranteed by previous studies~\cite{HR,DMZ,DSWZ} of the non-linear model's DoF count, it   verifies  the linearized mGR study~\cite{Bernard}. In that work, the fiducial metric is eliminated in terms of the mean field in order to argue that linearized spin~2 fields can propagate consistently in {\it any} gravitational background. This result is consistent with earlier work in~\cite{Gitman} which relies on  a $1/m^2$ expansion to study leading DoF and causality properties of gravitating, massive spin~2 models.} DoF.

 \section{Characteristic matrix}
We now study whether  \mGR can propagate initial data
off a given hypersurface~$\Sigma$. 
This amounts to asking if derivatives normal to~$\Sigma$ are determined by the equations of motion~\eqref{mbGR}. For that, we simply replace all derivatives in the equations of motion and gradients of their constraints by the normal covector~$\xi_\mu$  to $\Sigma$ multiplied by the normal derivative of the corresponding field:
$$
\partial_\mu \varepsilon^m\big|_\Sigma = \xi_\mu \partial_{\rm n}
\varepsilon^m
\ \mbox{ and }\ 
\partial_\mu \lambda^{mn}\big|_\Sigma = \xi_\mu \partial_{\rm n}\lambda^{mn} \, .
$$
We will also, for reasons of simplicity alone,
restrict to the parameter choices $\beta_2=\beta_3=0$ 
(the model's characteristic matrix for its entire
parameter range has been computed in~\cite{DSWZ}). In particular we must focus on the question whether the linear  system of equations for the normal derivatives~${\partial_{\rm n} \varphi}:= ({{\partial_{\rm n} \varepsilon},{\partial_{\rm n} \lambda}})$ implied by the equations of motion along~$\Sigma$ is invertible. This amounts to a matrix problem encoded by the theory's characteristic matrix~${\mathcal C}$. In what follows we compute the system of equations given by the homogeneous linear system ${\mathcal C} \cdot {\partial_{\rm n} \varphi} = 0$.
Starting with the equations of motion we find
\begin{eqnarray}\label{cheom}
&\xi\wedge  \partial_{\rm n} \varepsilon^m =  0 \, , \\[1mm]
&\epsilon_{mnrs} e^n\wedge \xi \wedge \partial_{\rm n} \lambda^{rs}=0\, .\nn
\end{eqnarray}
The gradients of the secondary and tertiary constraints then imply 
\begin{eqnarray}\label{2and3}
&\xi_\mu\,  f_m \wedge \partial_{\rm n} \varepsilon^m  = 0 = 
\xi_{\mu}\,  \epsilon_{mnrs} e^m \wedge \big[e^n \wedge \partial_{\rm n} \lambda^{rs}-2   K^{nr}\wedge \partial_{\rm n} \varepsilon^s\big]\, ,
&\\[1mm]
&\xi_\mu\,  f^m\wedge \big[
K_{mn} \wedge \partial_{\rm n}\varepsilon^n +
e^n \wedge \partial_{\rm n} \lambda_{mn}
\big]=0\, ,&
\nn
\\[1mm]
&\xi_\mu\, 
\epsilon_{mnrs}e^{m}\!\wedge\!\big[e^{n}\!\wedge\! K^{rt}\!\wedge\!\partial_{\rm n}\lambda_{t}{}^{s}+\big(K^{nt}\!\wedge\! K_{t}{}^{r}-\bar{R}^{nr}+m^{2}(4\beta_{0}e^{n}\!\wedge\! e^{r}+3\beta_{1}e^{n}\!\wedge\! f^{r})\big)\!\wedge\!\partial_{\rm n}\varepsilon^{s}\big] = 0 \ .
\!\!\!\!\!\!\!&
\nn
\end{eqnarray}
In the above the prefactor $\xi_\mu$ was included to indicate the origin of these equations but can be removed with impunity. To handle Equation~\eqref{cheom} we decompose form-valued normal derivatives as earlier in Equation~\eqref{ring}, 
and find\footnote{\label{fineprint} Here we assumed that the pullback of the mGR background vierbeine to the hypersurface~$\Sigma$ is invertible and ignore likely  pathologies when this fails.} 
$$
\partial_{\rm n} {\bm \varepsilon}^{m}=0=\partial_{\rm n} {\bm \lambda}^{mn}\, .
$$
Supposing that the one-form $\xi = dt$, for some evolution coordinate $t$, we now use a shorthand notation $\partial_n \varepsilon^{m} = dt \, \dot \varepsilon^m_t$ and $\partial_n \lambda^{mn}= dt \, \dot \lambda^{mn}_t$.
We thus have the {\it reduced characteristic system} 
\begin{equation}\label{reduce}
\begin{pmatrix}
{\bm f}_m & 0\\[1mm]
2\epsilon_{mnrs} {\bm e}^n\times {\bm K}^{rs}&
\epsilon_{mnrs} {\bm e}^r\times {\bm e}^s\\[1mm]
{\bm f}^n\times {\bm K}_{nm}&
{\bm f}_m\times {\bm e}_n\\[1mm]
{\mathcal R}_m&{\mathcal K}_{mn}
\end{pmatrix}
\begin{pmatrix}
\dot\varepsilon_t^m \\[2mm]
\dot\lambda_t^{mn}
\end{pmatrix}=0\, .
\end{equation}
In the above square matrix, $\times$ denotes the standard three-dimensional cross product while the spatial densities on its last line can be read off from~\eqref{2and3} and 
are  simple for  flat fiducial metrics. 
Vanishing of the determinant of the above $10\times 10$ matrix completely characterizes the boundary of the model's predictive hyperbolic regime (modulo the restriction explained in Footnote~\ref{fineprint}). 
As we shall see, the reduced characteristic system describes the propagation of superluminal lower helicity modes:
In the next section, we specialize to flat fiducial spaces and show how to analyze this determinant in direct analogy with the RS system.

\section{Analogy with Rarita--Schwinger}

The charged, spin~$3/2$, Rarita--Schwinger (RS) equation of motion reads
$$
\gamma^{\mu\nu\rho} \big(\nabla_\nu + ie A_\nu + \frac m2\,  \gamma_\nu\big)\psi_\rho=0\, .
$$
Here $\nabla$ is the Levi-Civita connection of
the fiducial spacetime and $A$ is the background EM potential. These are analogous to the  \mGR fiducial and background-mean-field metrics. The RS characteristic matrix was computed in~\cite{VZ} for flat fiducial metrics and spacelike hypersurfaces $\Sigma$ and found to have zero determinant when the magnetic field $\bm B$ obeyed\footnote{See~\cite{Porrati} for models designed to cure this pathology by adding higher background derivative, string-inspired terms to the RS action.}
\begin{equation}\label{R}
1-\Big(\frac{2e}{3m^2}\Big)^{\!2} {\bm B}^2 =0\, .
\end{equation}
The condition $\frac{2e|B|}{3m^2}<1$ thus determines the weak field, hyperbolic, regime.
Our aim now is to develop the analogous statement for \mGR and inherit the conclusions of~\cite{VZ}.

We begin with a short calculation.
Consider  now  a flat fiducial metric so $f^m=\delta^m_\mu dx^\mu$ and
$d\bar s^2=-dt^2 +  d{\bm x}^{ 2}$. This simplifies the reduced characteristic system considerably.
Firstly the equation ${\bm f}_m \dot\varepsilon^m_t=0$ of~\eqref{reduce} implies $\dot\varepsilon^a_t=0$, where we have decomposed the Lorentz index $m=(0,a)$. Let us introduce an EM-like notation
$$
\epsilon_{abc} {\bm K}^{bc}=:{\bm B}_a\, ,\quad 
{\bm K}^{0a}=:{\bm E}^a
\, .
$$
For pure simplicity reasons only, we now restrict to the case where $\bm e^{0}=\bm E^{a}=\bm 0\ .$
Thus using $\epsilon^0{}_{abc}=\epsilon_{abc}$ the second equation of~\eqref{reduce} implies
$$
\dot\lambda_t^{0a}=-\frac12 \epsilon^{abc} {\bm B}_b\cdot\tilde {\bm e}_c \, \dot\varepsilon_t^0
\, .
$$
Here the 3-vectors $\tilde {\bm e}_a$ form the 3-inverse of ${\bm e}^a$ so that $\tilde {\bm e}_a\cdot {\bm e}^b=\delta^b_a$.
The third equation of~\eqref{reduce} then gives
$$
\bm f_{[a}\times\bm e_{b]}
\dot\lambda_{t}^{ab}=0\ .
$$
This equation generically allows  $\dot\lambda_{t}^{ab}$ to be expressed as a function of $\dot\varepsilon_{t}^{0}$ but  this requires a non-trivial  condition on $\bm e^{a}$.  Under the hypothesis that the eigenvalue spectra of  the matrices $\bm f^{a}\cdot\tilde{\bm e}_{b}$ and $-\bm f^{a}\cdot\tilde{\bm e}_{b}$ do not intersect, the above equation implies that $\dot\lambda_{t}^{ab}=0$. In this framework, the last equation of~\eqref{reduce} gives the single VZ-type condition 
\begin{eqnarray}\label{m}
\left[m_{\text{FP}}^{2}\big(4-(\bm f^{a}\cdot\tilde{\bm e}_{a})\big)-\frac16\Big( (\bm B^{a}\cdot\tilde{\bm e}^{b})(\bm B_{[a}\cdot\tilde{\bm e}_{b]})-\frac12(\bm B^{a}\cdot\tilde{\bm e}_{[a})(\bm B^{b}\cdot\tilde{\bm e}_{b]})\Big)\right]\dot\varepsilon_{t}^{0}=0\ .
\end{eqnarray}
In the weak field limit where the mean field approaches
Minkowski space, the coefficient of~$m_{\rm FP}^2$ approaches unity but can change sign in a strong-field, large $\tilde {\bm e}$ limit. Hence there are certainly strong field configurations where the model loses hyperbolicity~\cite{CourantHilbert} and closed causal curves are unavoidable.  
(This signals the onset of strong coupling in an effective field theory.)
Now, comparing Equations~\eqref{R} and~\eqref{m}, we see that we have reduced \mGR\!\!'s 
weak field propagation analysis to a previous---well understood---case. Finally, we note that the same analogy and characteristic method can also be applied to the  bimetric theory by treating the two background metrics as a (fiducial,background) pair~\cite{prep}.

\section{Conclusions}

mGR is not a fundamental theory but rather an effective one  with a range of validity determined by requiring  hyperbolicity  in a weak field regime. Excepting the further caveats explained in the text, the reduced characteristic matrix of Equation~\eqref{reduce} completely determines this allowed regime. Even in the weak regime, modes exhibit a crystal structure with differing maximal propagation speeds.
For mGR to give a useful effective theory for
physical applications,  one must couple to matter (or at least photons) and require consistent 
causal cones for all modes. Once these couplings  are decided upon, the characteristic method will determine their effective range of validity, if any. 
There is also the logical possibility that there exists a causal, luminal, UV completion of mGR analogous to that for QED in curved space~\cite{Hollowood} (see however the more general discussion of UV completions~\cite{Dubovsky}).

\section*{Acknowledgements}
We thank C. Deffayet, S. Dubovsky, K. Hinterbichler, K. Izumi, M. Porrati and Y.C. Ong  for discussions. A.W. and G.Z. thank the Perimeter Institute for an illuminating 
``Superluminality in Effective Field Theories for Cosmology''
workshop.
S.D. was supported in part by grants NSF PHY-1266107 and DOE \# de-sc0011632. A.W.
was supported in part by a Simons Foundation Collaboration Grant for Mathematicians. G.Z. was supported in part by DOE Grant DE-FG03-91ER40674.

\end{document}